\documentclass[a4paper,twocolumn,journal]{IEEEtran}
\IEEEoverridecommandlockouts

\usepackage{amsfonts, amssymb, amsmath, cite, enumerate, amsthm, bm, caption, subfig, psfrag, cases, mathtools}
\usepackage{ifthen}
\usepackage[usenames,dvipsnames]{color}
\usepackage{epstopdf}
\usepackage{graphicx}

\DeclareMathOperator{\E}{\mathbb{E}}

\DeclareMathOperator{\dist}{dist}

\DeclareMathOperator{\sign}{sign}

\newcommand{\ip}[2]{\left\langle#1,#2\right\rangle}
\newcommand{\norm}[1]{\left\lVert#1\right\rVert}
\renewcommand{\(}{\left(}
\renewcommand{\)}{\right)}
\renewcommand{\[}{\left[}
\renewcommand{\]}{\right]}

\newboolean{draft}
\newcommand{\isdraft}[2]{\ifthenelse{\boolean{draft}}{#1}{#2}}

\isdraft{\usepackage{setspace}}{}                              % DRAFT
\isdraft{\usepackage[footnotesize]{caption}}{}                 % DRAFT
\isdraft{\usepackage{paralist}}{}                              % DRAFT

\def \R {\mathbb{R}}

\def \< {\langle}
\def \> {\rangle}

\def \vd {\bm{d}}

\def \vg {\bm{g}}
\def \vh {\bm{h}}
\def \vn {\bm{n}}
\def \vp {\bm{p}}

\def \vx {\bm{x}}

\def \vw {\bm{w}}
\def \vy {\bm{y}}

\def \vA {\bm{A}}

\def \vphi {\bm{\phi}}

\theoremstyle{plain}
\newtheorem{theorem}{Theorem}
\newtheorem{proposition}{Proposition}

\newtheorem{lemma}{Lemma}

\theoremstyle{remark}
\newtheorem{remark}{Remark}

\begin{document}
%\setboolean{draft}{true}
\title{Compressed Sensing with Prior Information via Maximizing Correlation}

\author{
\IEEEauthorblockN{Xu~Zhang\IEEEauthorrefmark{1}, Wei~Cui\IEEEauthorrefmark{1}, and Yulong~Liu\IEEEauthorrefmark{2}}

\IEEEauthorblockA{\IEEEauthorrefmark{1}School of Information and Electronics, Beijing Institute of Technology, Beijing 100081, China}

\IEEEauthorblockA{\IEEEauthorrefmark{2}School of Physics, Beijing Institute of Technology, Beijing 100081, China}

\thanks{This work was supported by the National Natural Science Foundation of China under Grants 61301188 and 61672097.}
}
\date{}

%\markboth{Journal of \LaTeX\ Class Files,~Vol.~XX, No.~X, Month~Year}%
%{Compressed Sensing with Prior Information via Maximizing Correlation}



\maketitle

\pagestyle{empty}  % no page number for the second and the later pages
\thispagestyle{empty} % no page number for the first page

\begin{abstract}
Compressed sensing (CS) with prior information concerns the problem of reconstructing a sparse signal with the aid of a similar signal which is known beforehand. We consider a new approach to integrate the prior information into CS via maximizing the correlation between the prior knowledge and the desired signal. We then present a geometric analysis for the proposed method under sub-Gaussian measurements. Our results reveal that if the prior information is good enough, then the proposed approach can improve the performance of the standard CS. Simulations are provided to verify our results.
\end{abstract}


% \fi
\IEEEpeerreviewmaketitle

\section{Introduction}
Compressed sensing concerns the problem of recovering a high-dimensional sparse (or nearly sparse) signal from a relatively small number of noisy measurements
\begin{equation}\label{Observation_model}
  \vy= \vA \vx^\star + \vn,
\end{equation}
where $\vA \in \R^{m \times n}$ is the measurement matrix, $\vx^\star \in \R^n$ denotes the signal to be estimated, and $\vn \in \R^m$ is the observation noise vector. See e.g., \cite{donoho2006compressed,candes2006robust,Candes2006Near}. To estimate $\vx^\star$, a standard approach, named Lasso \cite{tibshirani1996regression} or Basis Pursuit \cite{chen2001atomic}, was proposed to use $\ell_1$-norm as a surrogate function to promote sparsity, that is
\begin{equation}\label{eq: Classical_Problem}
    \min \limits_{\vx \in \R^n} \norm{\vx}_1 ~~~~\text{s.t.} ~~ \|\vy-\vA \vx\|_2 \leq \delta,
\end{equation}
where $\|\vx\|_p = \left(\sum_{i=1}^{n}|\vx_i|^p\right)^{1/p}$ denotes the standard $\ell_p$-norm of $\vx$, and $\delta$ is the upper bound (in terms of $\ell_2$ norm) of the noise vector $\vn$.

However, in many practical applications, it is possible to have access to some prior knowledge about the desired signal in addition to the sparsity constraint. For instance, in video acquisition \cite{stankovi2009compressive,kang2009distributed} and dynamic system estimation \cite{charles2011sparsity}, past signals are very similar to the signal to be acquired, and hence they can be utilized as prior information. Then it is highly desirable to employ these prior knowledge to improve the performance of the standard CS.
There are a number of forms in the literature to integrate prior information into standard CS, see, e.g., \cite{chen2008prior,vaswani2010modified,charles2011sparsity,khajehnejad2011analyzing,scarlett2013compressed,mota2017compressed} and references therein.
A class of methods is to modify the objective in \eqref{eq: Classical_Problem} by adding a new penalty term which penalizes the differences between $\vx$ and the prior information $\vphi$, that is
\begin{equation}\label{eq: Original_Problem}
    \min \limits_{\vx} \norm{\vx}_1+ \lambda g(\vx-\vphi) ~~~~\text{s.t.} ~~  \|\vy-\vA \vx\|_2 \leq \delta,
\end{equation}
where $\lambda > 0$ establishes a tradeoff between signal sparsity and fidelity to prior information, and $g: \R^n \rightarrow \R$ is a function which measures the similarity between $\vx$ and $\vphi$. Examples of $g$ include $g(\vx-\vphi)=\norm{\vx-\vphi}_1$ and $g(\vx-\vphi)=\frac{1}{2} \norm{\vx-\vphi}_2^2$, which is called $\ell_1$-$\ell_1$ minimization and $\ell_1$-$\ell_2$ minimization respectively.

In this paper, we consider a new approach to incorporate prior information into Lasso by maximizing the correlation between $\vx$ and $\vphi$, which leads to
\begin{equation}\label{eq: Main_Problem}
    \min \limits_{\vx} \norm{\vx}_1 - \lambda \langle \vx, \vphi\rangle ~~~~\text{s.t.} ~~  \|\vy-\vA \vx\|_2 \leq \delta,
\end{equation}
where $\lambda > 0$ is a tradeoff parameter. This is motivated by the observation that if $\vphi$ is very similar to $\vx^\star$, then they may be highly correlated. We also present a geometric analysis for the proposed method under sub-Gaussian measurements. Specifically, we show that if the prior information $\vphi$ is good enough, then \eqref{eq: Main_Problem} can improve the performance of the standard CS.
%
%
%We make three contributions in this paper:
%\begin{enumerate}
%  \item We propose a new approach, that is MaxCorrealtion, for CS with prior information;
%  \item We establish the CS bound for convex function with or without noise under sub-Gaussian measurements;
%  \item We present the theoretical analysis of MaxCorrealtion and provide the bound on the number of measurements required to recover original signal in high probability.
%\end{enumerate}

\section{Performance guarantees}
In this section, we begin with introducing some preliminaries which will be used in this paper, and then establish the performance guarantees for the proposed approach in a geometric way.

\subsection{Preliminaries}
A random variable $x$ is a \emph{sub-Gaussian random variable} if it has finite Orlicz norm $\norm{x}_{\psi_2}$, defined by
$$
\norm{x}_{\psi_2}=\inf \{t>0: \E \exp(x^2/t^2)\le 2\}.
$$
The sub-Gaussian norm of $x$, denoted $\norm{x}_{\psi_2}$, is the smallest $t$ such that $\E \exp(x^2/t^2)\le 2$. A random vector $\vx \in \R^n$ is called a \emph{sub-Gaussian random vector} if all of its one-dimensional marginals are sub-Gaussian, and its sub-Gaussian norm is defined as
$$
\norm{\vx}_{\psi_2} = \sup \limits_{\vy \in S^{n-1}} \norm{\left\langle \vx,\vy \right\rangle}_{\psi_2}.
$$
We call a random vector $\vx \in \R^n$  \emph{isotropic} if it satisfies $\E \vx \vx^T = \bm{I}_n$, where $\bm{I}_n$ is the $n$-dimensional identity matrix.

The \emph{tangent cone} of a convex function $f: \R^n \to \R$ at $\vx^\star$ is defined as
$$
    \mathcal{T}_f = \{\vd \in \R^n: f(\vx^\star+t \cdot \vd) \le f(\vx^\star)~~\textrm{for some}~t>0\},
$$
which is the set of descent directions of $f$ at $\vx^\star$.
The \emph{normal cone} of  a convex function $f$ at $\vx^\star$ is the polar cone of the tangent cone
$$
\mathcal{N}_f= \{\vp \in \R^n: \ip{\vd}{\vp} \le 0 ~~ \textrm{for all}~\vd \in \mathcal{T}_f\}.
$$

The \emph{Gaussian width} and  \emph{Gaussian complexity} of a subset $\mathcal{E} \subset \R^n$ are defined as
$$
    w(\mathcal{E})= \E \sup \limits_{\vx \in \mathcal{E}} \ip{\vg}{\vx},~\vg \sim N(0,\bm{I}_n)
$$
and
$$
    \gamma (\mathcal{E})= \E \sup \limits_{\vx \in \mathcal{E}} |\ip{\vg}{\vx}|,~\vg \sim N(0,\bm{I}_n)
$$
respectively.
These two geometric quantities have a close relationship, that is,
\begin{equation}\label{Relation}
\gamma(\mathcal{E}) \le  2 w(\mathcal{E})+\norm{\vy}_2~~~\textrm{for any point}~~~\vy \in \mathcal{E}.
\end{equation}

We also use the following matrix deviation inequality which implies the restricted eigenvalue condition for the sub-Gaussian sensing matrix.
\begin{proposition}[Matrix deviation inequality, \cite{liaw2016simple}] \label{lm:MatrixDeviationInequality}
 Let $\vA$ be an ${m \times n}$ random matrix whose rows are independent, centered, isotropic and sub-Gaussian random vectors. For any bounded subset $\mathcal{D} \subset \R^n$ and $t \ge 0$, the event
$$
    \sup \limits_{\vx \in \mathcal{D}} \left| \norm{\vA \vx}_2 - \sqrt{m} \norm{\vx}_2 \right| \le CK^2 [\gamma (\mathcal{D})+t\cdot \text{\emph{rad}}(\mathcal{D})]
$$
holds with probability at least $1-2\exp(-t^2)$. Here $K=\max_i \norm{\vA_i}_{\psi_2}$ and $ \text{\emph{rad}}(\mathcal{D})=\sup_{\vx \in \mathcal{D}} \norm{\vx}_2$.
\end{proposition}

\subsection{Main results}

In this subsection, we establish theoretical results for the proposed approach \eqref{eq: Main_Problem}. Our main results show that if the prior information is good enough, the proposed approach can achieve a better performance than that of Lasso \eqref{eq: Classical_Problem}.

\begin{theorem} \label{thm: Generalcase} Let $\vA \in \R^{m \times n}$ be a random matrix whose rows are independent, centered, isotropic and sub-Gaussian random vectors, and let $\mathcal{T}_f$ denote the tangent cone of $f(\vx):=\norm{\vx}_1 - \lambda \ip{\vphi}{\vx}$ at $\vx^\star$. If
\begin{equation}\label{NumberofMeasurements}
        \sqrt{m} \ge CK^2 \gamma(\mathcal{T}_f \cap S^{n-1})+ \epsilon,
\end{equation}
    then with probability at least $1- 2\exp(-\gamma^2(\mathcal{T}_f \cap S^{n-1}))$, the solution $\hat{\vx}$ to \eqref{eq: Main_Problem} satisfies
    $$
    \norm{\hat{\vx}-\vx^\star}_2 \le \frac{2\delta}{\epsilon},
    $$
    where $\epsilon,C$ are absolute constants and $K=\max_i \norm{\vA_i}_{\psi_2}$.
\end{theorem}
\begin{IEEEproof}
Let $\vh = \hat{\vx} - \vx^\star$. Since $\hat{\vx}$ solves \eqref{eq: Main_Problem}, we have $\vh \in \mathcal{T}_f$ and $\vh/\norm{\vh}_2 \in \mathcal{T}_f \cap S^{n-1}$. It follows from Proposition \ref{lm:MatrixDeviationInequality} (let $\mathcal{D} = \mathcal{T}_f \cap S^{n-1}$) that the event
\isdraft{
\begin{equation}\label{Restricted_Eigenvulue}
   \sqrt{m}- \inf \limits_{\bar{\vh} \in \mathcal{T}_f \cap S^{n-1}} \norm{\vA \bar{\vh}}_2\leq \sup \limits_{\bar{\vh} \in \mathcal{T}_f \cap S^{n-1}} \left| \norm{\vA \bar{\vh}}_2 - \sqrt{m} \right| \le C'K^2 [\gamma(\mathcal{T}_f \cap S^{n-1})+t]
\end{equation}
}
{
\begin{equation}\label{Restricted_Eigenvulue}
 \begin{aligned}
   \sqrt{m}- \inf \limits_{\bar{\vh} \in \mathcal{T}_f \cap S^{n-1}} \norm{\vA \bar{\vh}}_2 &\leq \sup \limits_{\bar{\vh} \in \mathcal{T}_f \cap S^{n-1}} \left| \norm{\vA \bar{\vh}}_2 - \sqrt{m} \right| \\
    &\le C'K^2 [\gamma(\mathcal{T}_f \cap S^{n-1})+t]
   \end{aligned}
\end{equation}
}
holds with probability at least $1- 2\exp(-t^2)$. Here we have used the facts that $\norm{\bar{\vh}}_2=1$ and $\text{rad}(\mathcal{T}_f\cap S^{n-1})=1$. Choose $t = \gamma(\mathcal{T}_f \cap S^{n-1})$ in \eqref{Restricted_Eigenvulue} and if $\sqrt{m} \ge CK^2 \gamma(\mathcal{T}_f \cap S^{n-1})+ \epsilon$, we have the event
\begin{align*}
 \inf \limits_{\bar{\vh} \in \mathcal{T}_f \cap S^{n-1}}\norm{\vA \bar{\vh}}_2\geq  \sqrt{m}-2C'K^2\gamma (\mathcal{T}_f \cap S^{n-1})\geq \epsilon
 \end{align*}
holds with probability at least $1- 2\exp(-\gamma^2(\mathcal{T}_f \cap S^{n-1}))$, where $C = 2C'$. Therefore, with desired probability, we have
\begin{equation}\label{eq: 2.2}
\|\vA\vh\|_2 = \norm{\hat{\vx}-\vx^\star}_2 \norm{\vA \frac{\vh}{\norm{\vh}_2}}_2 \geq \epsilon \norm{\hat{\vx}-\vx^\star}_2.
\end{equation}

On the other hand, since both $\hat{\vx}$ and $\vx^\star$ are feasible, by triangle inequality, we obtain
\begin{equation}\label{eq: 2.1}
\|\vA\vh\|_2 \leq \norm{\vy-\vA\vx^\star}_2+\norm{\vy-\vA\hat{\vx}}_2 \leq 2 \delta.
\end{equation}

Combining \eqref{eq: 2.2} and \eqref{eq: 2.1} completes the proof.

\end{IEEEproof}

\begin{remark} Actually, Theorem \ref{thm: Generalcase} holds for any convex function. This theorem extends the result in \cite[Corollary 3.3]{chandrasekaran2012convex} from Gaussian measurements to sub-Gaussian measurements.
\end{remark}

\begin{remark} By the relationship between Gaussian width and Gaussian complexity \eqref{Relation}, the condition \eqref{NumberofMeasurements}
 can also be expressed in terms of Gaussian width $\sqrt{m} \ge CK^2 [2 \cdot w(\mathcal{T}_f \cap S^{n-1})+1]+ \epsilon = C'K^2 w(\mathcal{T}_f \cap S^{n-1})+ \epsilon $. The second inequality holds because in practical applications we usually have $w(\mathcal{T}_f \cap S^{n-1})>0$.
\end{remark}

\begin{remark} In the noiseless setting where $\delta=0$, Theorem \ref{thm: Generalcase} entails exact recovery of $\vx^\star$ with probability at least $1- 2\exp(-\gamma^2(\mathcal{T}_f \cap S^{n-1}))$ as long as $m \ge CK^4 \gamma^2(\mathcal{T}_f \cap S^{n-1})$.
\end{remark}

\begin{remark} The bounded noise in Theorem \ref{thm: Generalcase} can be easily extended to the sub-Gaussian case, since the sub-Gaussian distribution is bounded with high probability.
\end{remark}

%\begin{remark}
%From Theorem \ref{thm: Generalcase}, we can come to the conclusion that the number of measurements is determined by $w(\mathcal{T}_f_f \cap S^{n-1})$. Based on the relationship of normal cone and tangent cone, we can study the normal cone instead. When  $w(\mathcal{N} \cap S^{n-1})$ decreases,  $w(\mathcal{T}_f \cap S^{n-1})$ and the number of measurements will increase. This idea will be used in the geometrical analysis after Theorem \ref{thm: SpecialCase}.
%\end{remark}

To obtain an interpretable sample size bound in terms of familiar parameters, it is necessary to bound $\gamma(\mathcal{T}_f \cap S^{n-1})$ or $w(\mathcal{T}_f \cap S^{n-1})$. To this end, define
\begin{equation*}
  v:=\max \limits_{\vw \in \partial{\norm{\vx^\star}_1}-\lambda \vphi} \norm{\vw}_2^2.
\end{equation*}
Let $I=\{i:\vx_i^\star \neq 0\}$ and $\mathcal{H}$ be the space of vectors whose support only on $I$. Then the subdifferential of $\norm{\vx^\star}_1$ is given by
\begin{equation*}
  \partial \norm{\vx^\star}_1=\sign(\vx^\star)+\left\{\bm{\theta} \in \mathcal{H}^{\perp}: \max \limits_{i \in I^c} \norm{\bm{\theta}_i}_2 \le 1\right\},
\end{equation*}
where   $\mathcal{H}^{\perp}$ is the  orthogonal complement of $  \mathcal{H}$. A standard calculation yields
\isdraft{  %----------------------------draft
\begin{equation}\label{parameter_v}
v:= \sum \limits_{i \in I} (\sign(x_i^*)-\lambda \vphi_i)^2 + \sum \limits_{i \in I^c J} (1-\lambda \vphi_i)^2 \\ + \sum \limits_{i \in I^cJ^c}(1+\lambda \vphi_i)^2,
\end{equation}
}          %----------------------------draft
{
\begin{multline} \label{parameter_v}
    v:= \sum \limits_{i \in I} (\sign(\vx_i^*)-\lambda \vphi_i)^2 + \sum \limits_{i \in I^c \cap J} (1-\lambda \vphi_i)^2 \\ + \sum \limits_{i \in I^c \cap J^c}(1+\lambda \vphi_i)^2,
\end{multline}
}
where $J=\{i:\vphi_i < 0\}$. Then we have the following result.

\begin{figure*}[!t]
\centering
\subfloat[$\lambda_a\vphi_a=\[0.5,0\]^T$]{\includegraphics[width=1.25in]{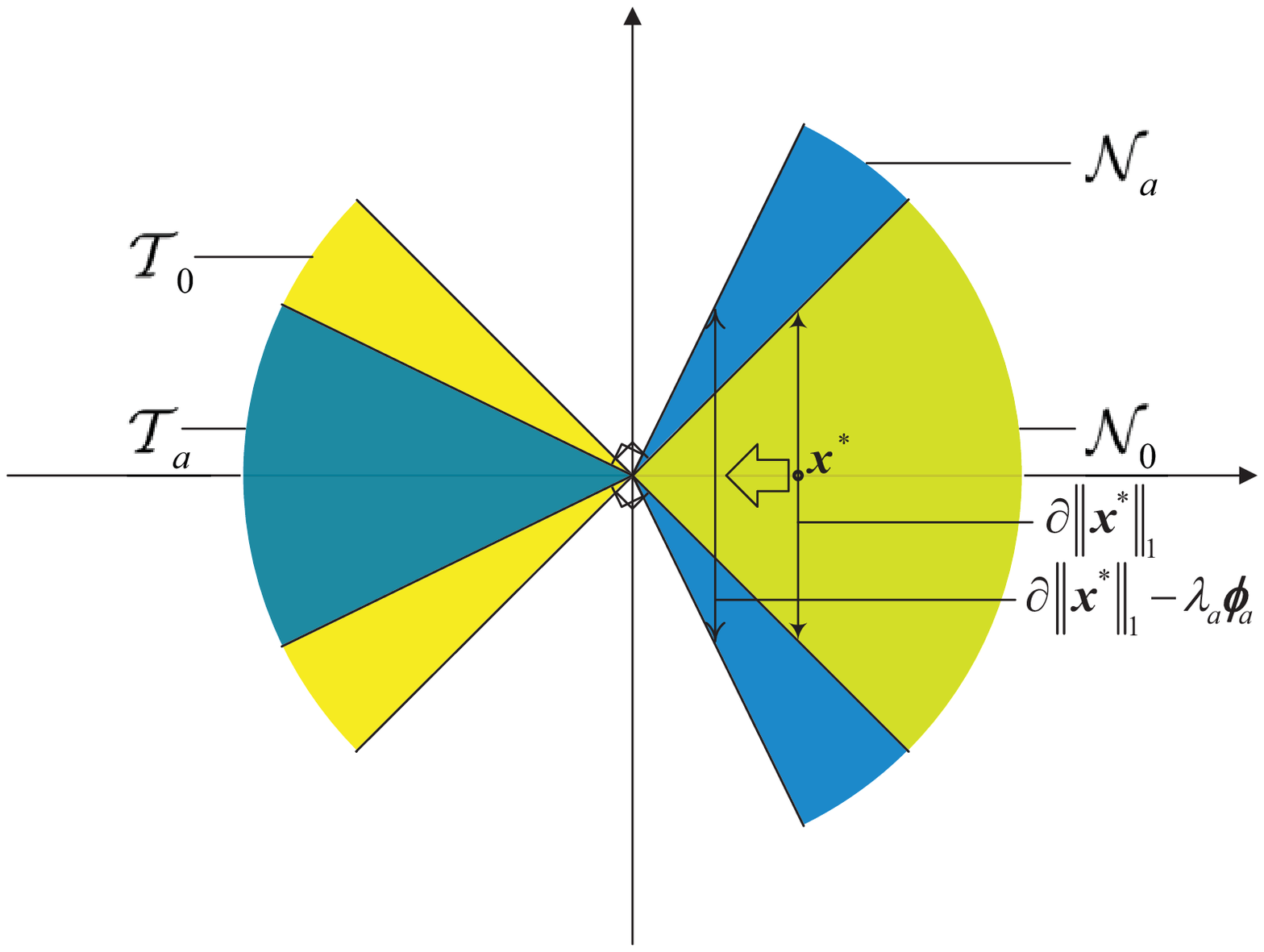}
\label{fig_first_case}}
\hfil
\subfloat[$\lambda_b\vphi_b=\[-0.5,0\]^T$]{\includegraphics[width=1.25in]{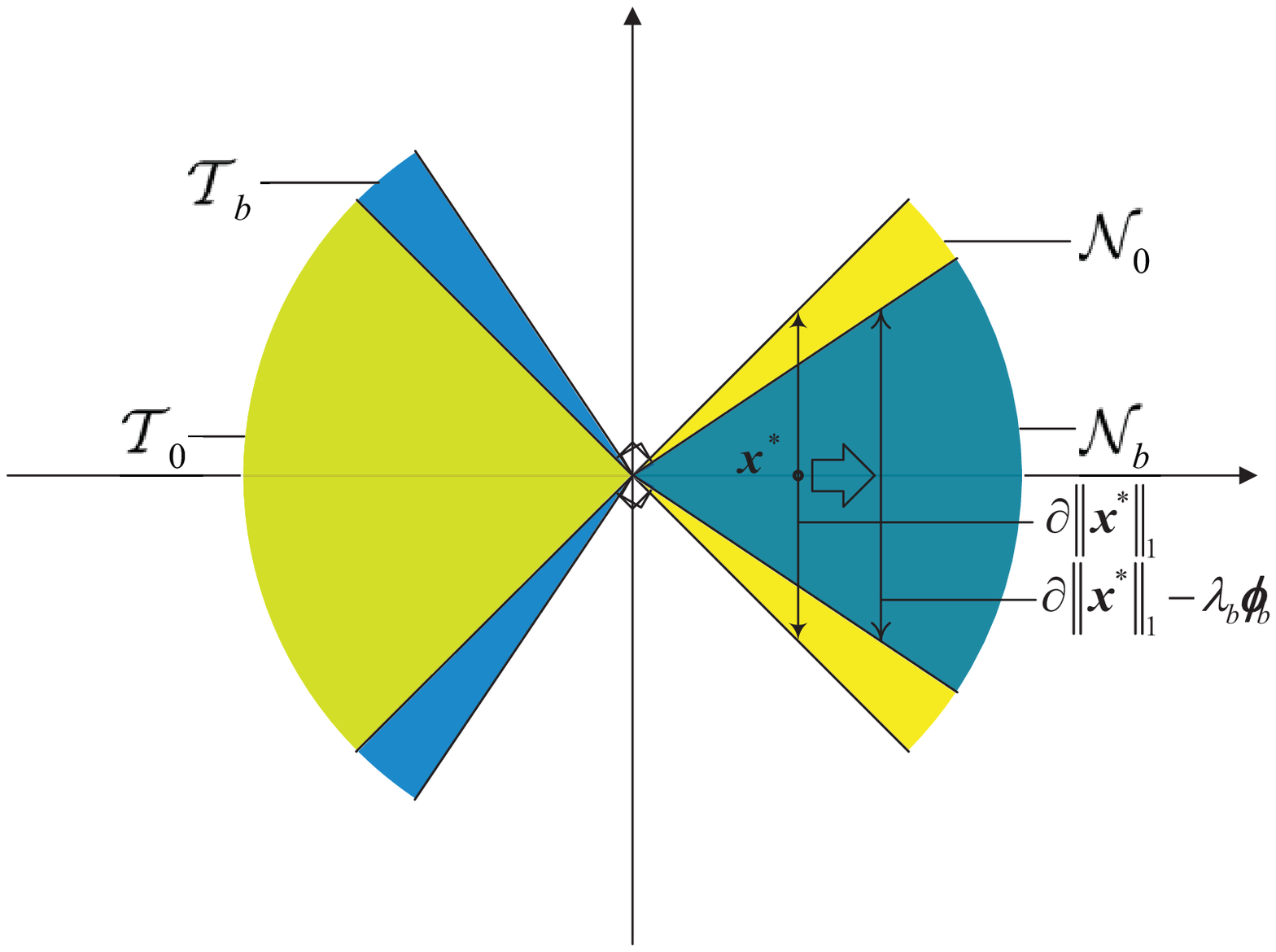}
\label{fig_second_case}}
\hfil
\subfloat[$\lambda_c\vphi_c=\[0,-1\]^T$]{\includegraphics[width=1.25in]{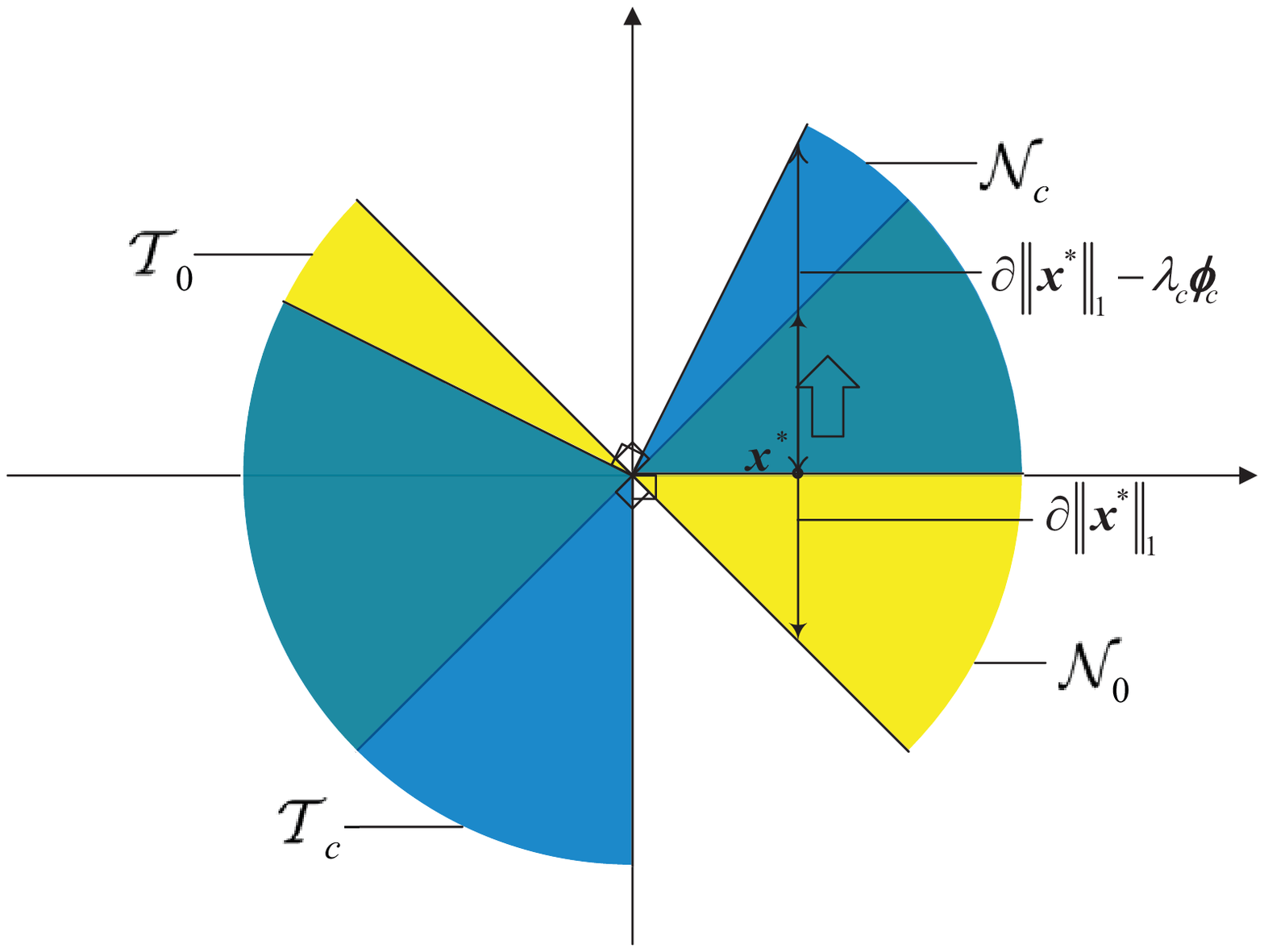}
\label{fig_third_case}}
\hfil
\subfloat[$\lambda_d\vphi_d=\[0.5,-0.2\]^T$]{\includegraphics[width=1.25in]{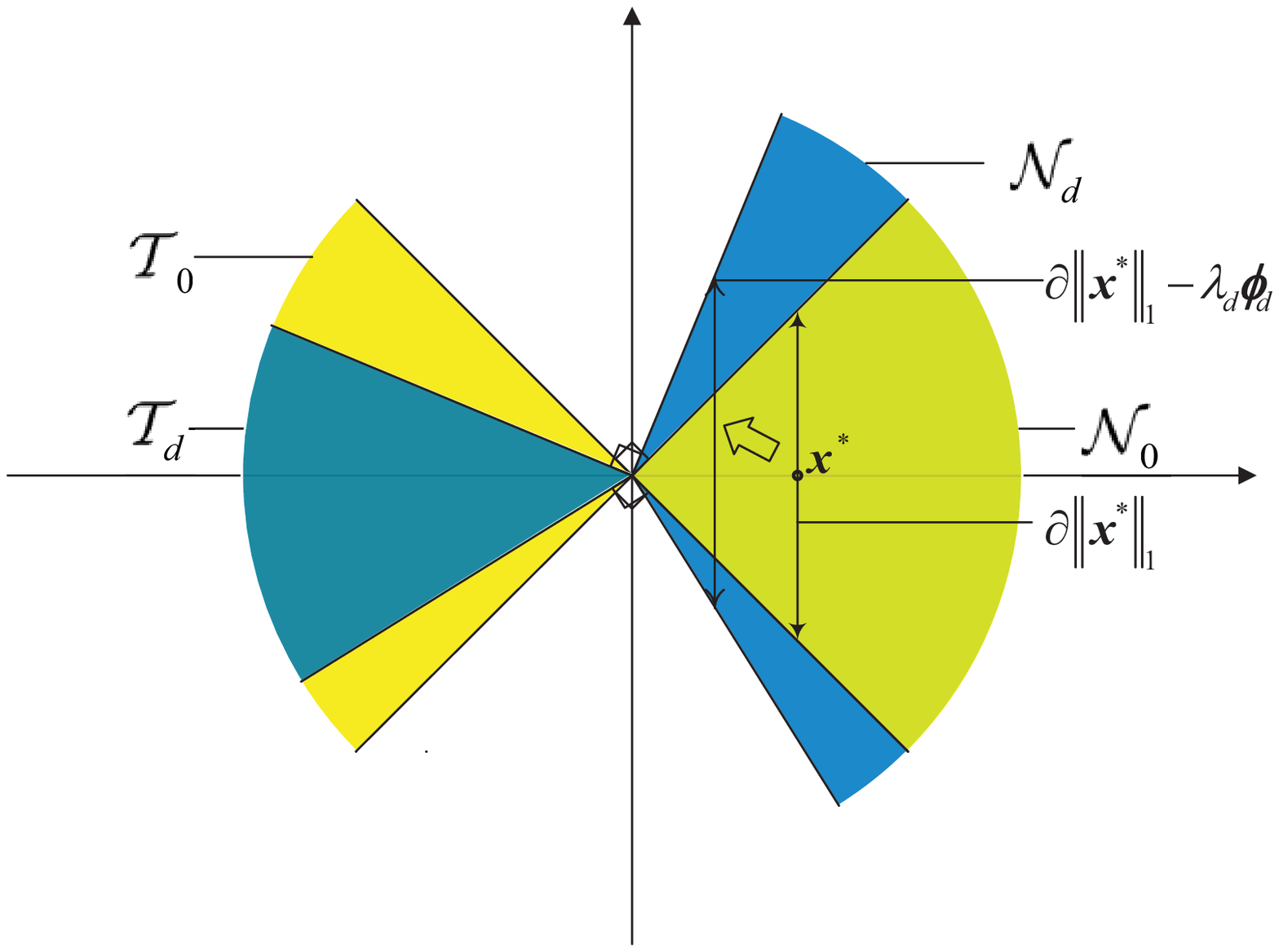}
\label{fig_fourth_case}}
\hfil
\subfloat[$\lambda_e\vphi_e=\[0.5,-1\]^T$]{\includegraphics[width=1.25in]{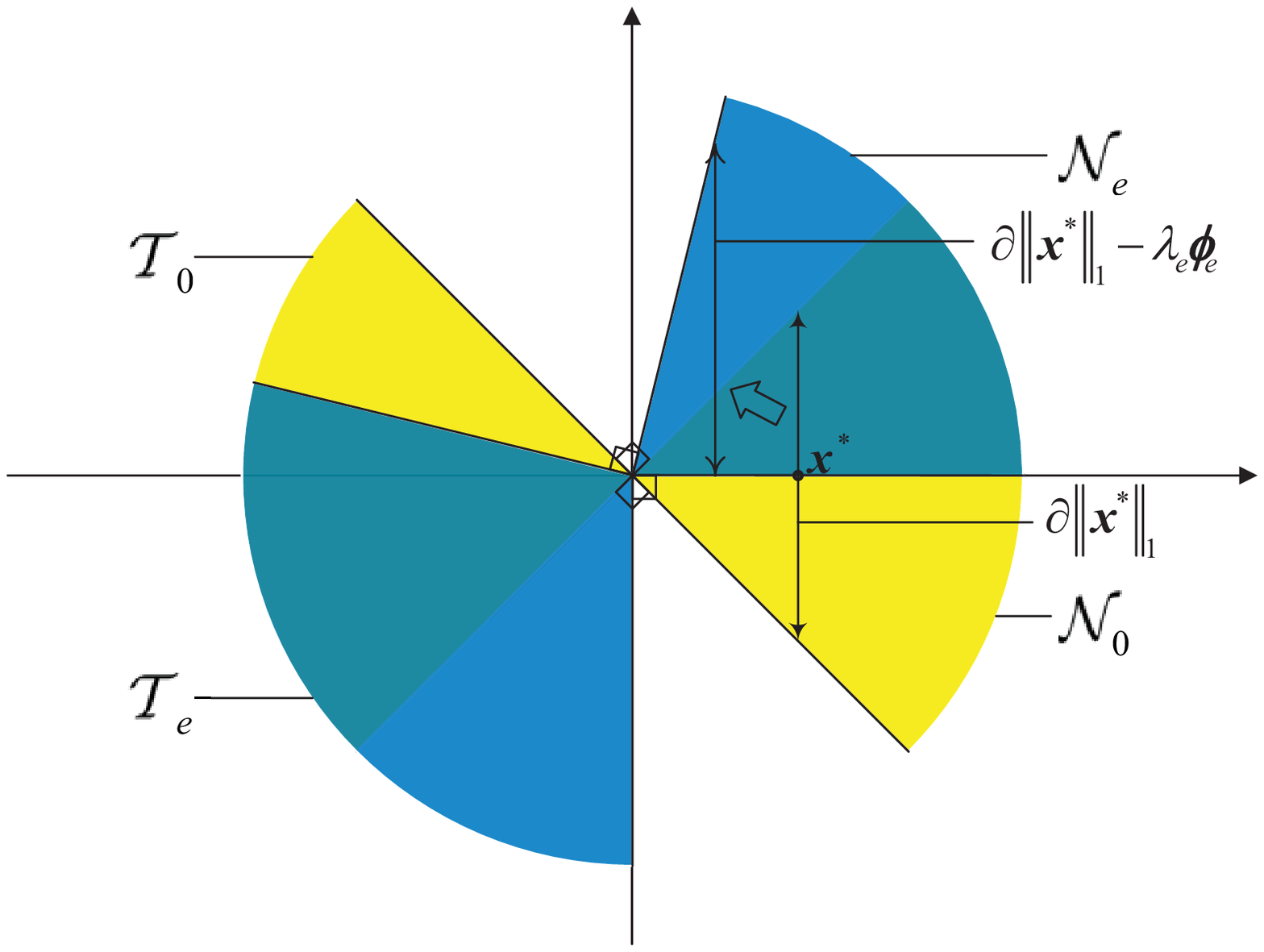}
\label{fig_fifth_case}}
\caption{The changes of normal cone and tangent cone after different shifts. For reference, we draw the tangent cone $\mathcal{T}_0$ and normal come $\mathcal{N}_0$ in yellow for classical CS in all figures. In each subfigure, the tangent cone $\mathcal{T}_\xi,~\xi=a,b,c,d,e$ and normal cone $\mathcal{N}_\xi,~\xi=a,b,c,d,e$ are shown in blue for the approach (\ref{eq: Main_Problem}) under different shifts. Good shifts, such as (a) and (d), enlarge the normal cone and hence decrease the number of measurements. Bad shifts, such as (b), (c), and (e), narrow the normal cone and lead to the growth of sample size.  }
\label{fig: GeometricalAnalysis}
\end{figure*}

\begin{lemma} \label{lm: CalculateGaussianWidth} Let $\vx^\star \in \R^n$ be an $s$-sparse vector and $\vphi \in \R^n$ be its prior information. Let $\mathcal{T}_f$ denote the tangent cone of $ f(\vx):=\norm{\vx}_1 - \lambda \ip{\vphi}{\vx}$ at $\vx^\star$. Suppose that $\bm{0} \notin \partial \norm{\vx^\star}_1 - \lambda \vphi$.
Then
$$
    w^2(\mathcal{T}_f \cap S^{n-1}) \le n \cdot \left( 1 - \frac{n}{v} \cdot \frac{2}{\pi} \(1-\frac{s}{n}\)^2 \right).
$$
\end{lemma}

\begin{IEEEproof}
According to \cite[Proposition 3.6]{chandrasekaran2012convex}, we have
\begin{equation*}
    w(\mathcal{T}_f \cap S^{n-1}) \le \E_{\vg} [\dist(\vg,\mathcal{N}_f)],
   % = \E_{\vg} [\dist(\vg, \text{\emph{cone}} \{\partial \norm{\vx^\star}_1 - \lambda \vphi\})]
\end{equation*}
where $\mathcal{N}_f$ denotes the polar of $\mathcal{T}_f$ and $\dist(\vx,\mathcal{E}):= \min \{ \norm{\vx-\vy}_2:\vy \in \mathcal{E}\}. $
Since $\bm{0} \notin \partial \norm{\vx^\star}_1 - \lambda \vphi$, it follows from \cite[Theorem 1.3.5]{hiriart1993convex} that
$$ \mathcal{N}_f =\text{cone} \{\partial \norm{\vx^\star}_1 - \lambda \vphi\}. $$
Thus, by Jensen's inequality, we have
$$
    w^2(\mathcal{T}_f \cap S^{n-1}) \le \E_{\vg} [\dist(\vg, \text{cone} \{\partial \norm{\vx^\star}_1 - \lambda \vphi\})^2].
$$

Fix any $\vg \in \R^n$ and choose any
$$\vw_0 \in \arg \max \limits_{\vw \in \partial \norm{\vx^\star}_1 - \lambda \vphi} \ip{\vg}{\vw},$$
then we obtain for any $t \geq 0$,
 \isdraft{\begin{equation*}   %----------------------------draft
    \begin{aligned}
 \dist(\vg, \text{\emph{cone}} \{\partial \norm{\vx^\star}_1 - \lambda \vphi\})^2 & \le \dist(\vg, t \cdot (\partial \norm{\vx^\star}_1 - \lambda \vphi))^2 \\
& \le \norm{\vg-t\vw_0}_2^2 =\norm{\vg}_2^2 -2t\ip{\vg}{\vw_0}+t^2\norm{\vw_0}_2^2\\
& = \norm{\vg}_2^2 -2t\max \limits_{\vw \in \partial \norm{\vx^\star}_1 - \lambda \vphi} \ip{\vg}{\vw}+t^2\norm{\vw_0}_2^2 \\
&\le \norm{\vg}_2^2 -2t\max \limits_{\vw \in \partial \norm{\vx^\star}_1 - \lambda \vphi} \ip{\vg}{\vw}+t^2 \max \limits_{\vw \in \partial \norm{\vx^\star}_1 - \lambda \vphi} \norm{\vw}_2^2.
   \end{aligned}
\end{equation*}              %-----------------------------draft
}{
\begin{equation*}
    \begin{aligned}
~& ~~~ \dist(\vg, \text{cone} \{\partial \norm{\vx^\star}_1 - \lambda \vphi\})^2 \\
 & \le \dist(\vg, t \cdot (\partial \norm{\vx^\star}_1 - \lambda \vphi))^2 &\\
& \le \norm{\vg-t\vw_0}_2^2 =\norm{\vg}_2^2 -2t\ip{\vg}{\vw_0}+t^2\norm{\vw_0}_2^2\\
& = \norm{\vg}_2^2 -2t\max \limits_{\vw \in \partial \norm{\vx^\star}_1 - \lambda \vphi} \ip{\vg}{\vw}+t^2\norm{\vw_0}_2^2\\
& \le \norm{\vg}_2^2 -2t\max \limits_{\vw \in \partial \norm{\vx^\star}_1 - \lambda \vphi} \ip{\vg}{\vw}+t^2 \max \limits_{\vw \in \partial \norm{\vx^\star}_1 - \lambda \vphi} \norm{\vw}_2^2.
   \end{aligned}
\end{equation*}
}
Taking expectation on both sides yields
\isdraft{                   %----------------------------draft
\begin{equation} \label{eq:: QuadraticBound}
    \begin{aligned}
\E\[ \dist(\vg, t \cdot (\partial \norm{\vx^\star}_1 - \lambda \vphi))^2\]
&\le n- 2t \cdot \E \max \limits_{\vw \in \partial \norm{\vx^\star}_1 - \lambda \vphi} \ip{\vg}{\vw} + t^2 \cdot \max \limits_{\vw \in \partial \norm{\vx^\star}_1 - \lambda \vphi} \norm{\vw}_2^2 \\
& = n- 2t \cdot \sqrt{\frac{2}{\pi}}(n-s) + t^2 \cdot v,
 \end{aligned}
\end{equation}
                            %----------------------------draft
}{
\begin{equation} \label{eq:: QuadraticBound}
    \begin{aligned}
   &\E_{\vg} \[ \dist(\vg, \text{cone} \{\partial \norm{\vx^\star}_1 - \lambda \vphi\})^2\] \\
&\le n- 2t \E \max \limits_{\vw \in \partial \norm{\vx^\star}_1 - \lambda \vphi} \ip{\vg}{\vw} + t^2  \max \limits_{\vw \in \partial \norm{\vx^\star}_1 - \lambda \vphi} \norm{\vw}_2^2 \\
& = n- 2t \sqrt{\frac{2}{\pi}}(n-s) + t^2 v,
 \end{aligned}
\end{equation}
}
where
$$
\E_{\vg} \max \limits_{\vw \in \partial \norm{\vx^\star}_1 - \lambda \vphi} \ip{\vg}{\vw}=\E_{\vg} \max \limits_{\vw \in \partial \norm{\vx^\star}_1} \ip{\vg}{\vw} =\sqrt{\frac{2}{\pi}}(n-s),
$$ and
\begin{equation*}
    v =\max \limits_{\vw \in \partial \norm{\vx^\star}_1 - \lambda \vphi} \norm{\vw}_2^2.
\end{equation*}
Choosing $t=\sqrt{2/ \pi} (n-s)/v$, we achieve the minimum of (\ref{eq:: QuadraticBound}). This completes the proof.
\end{IEEEproof}

\begin{remark} If there is no prior information, i.e., $\lambda \vphi=\bm{0}$, our approach (\ref{eq: Main_Problem}) reduces to the standard CS (\ref{eq: Classical_Problem}). In this case, \eqref{parameter_v} reduces to $v_0=n$ and hence
$$
 w^2(\mathcal{T}_0 \cap S^{n-1}) \le n \cdot \left( 1 - \frac{2}{\pi} \(1-\frac{s}{n}\)^2 \right),
$$
which coincides with the result in \cite[equation (9)]{foygel2014corrupted}. Here, $\mathcal{T}_0$ denotes the tangent cone of $\norm{\vx}_1$ at $\vx^\star$. However, if we choose some suitable $\lambda \vphi$ such that $v \le v_0=n$, then the approach (\ref{eq: Main_Problem}) can achieve better performance than the classical CS.
\end{remark}

Combining Theorem \ref{thm: Generalcase} with Lemma \ref{lm: CalculateGaussianWidth}, we arrive at the following result.

\begin{theorem} \label{thm: SpecialCase}Let $\vA$ be an ${m \times n}$ matrix whose rows are independent, centered, isotropic and sub-Gaussian random vectors and $\vx^\star \in \R^n$ be an $s$-sparse vector. Suppose that $ \bm{0} \notin  \partial \norm{\vx^\star}_1 -\lambda \vphi$. If
    $$
        \sqrt{m} \ge CK^2 \sqrt{ n \cdot \left( 1 - \frac{n}{v} \cdot \frac{2}{\pi} (1-\frac{s}{n})^2 \right)}+ \epsilon,
    $$
    then with probability $1- o(1)$, the solution $\hat{\vx}$ to (\ref{eq: Main_Problem}) satisfies
    $$
    \norm{\hat{\vx}-\vx^\star}_2 \le \frac{2\delta}{\epsilon},
    $$
    where $\epsilon,C$ are absolute constants and $K=\max_i \norm{\vA_i}_{\psi_2}$.
\end{theorem}

\begin{figure*}[!t]
\centering
\subfloat[]{\includegraphics[width=1.10in]{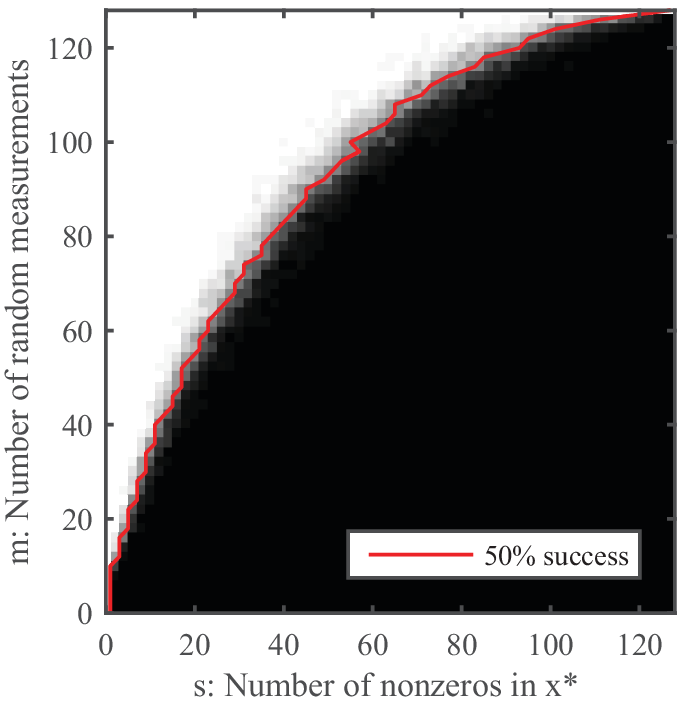}
\label{fig_first_case}}
\hfil
\subfloat[]{\includegraphics[width=1.10in]{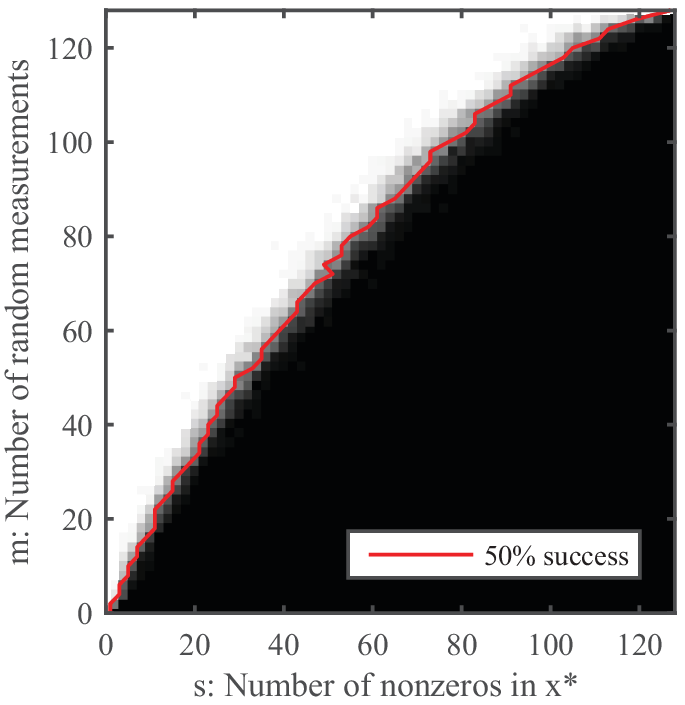}
\label{fig_second_case}}
\hfil
\subfloat[]{\includegraphics[width=1.10in]{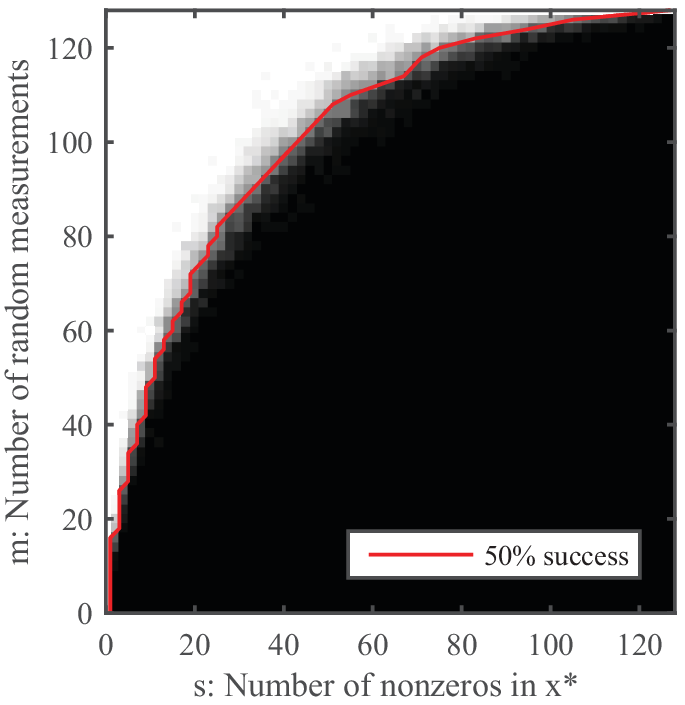}
\label{fig_third_case}}
\hfil
\subfloat[]{\includegraphics[width=1.10in]{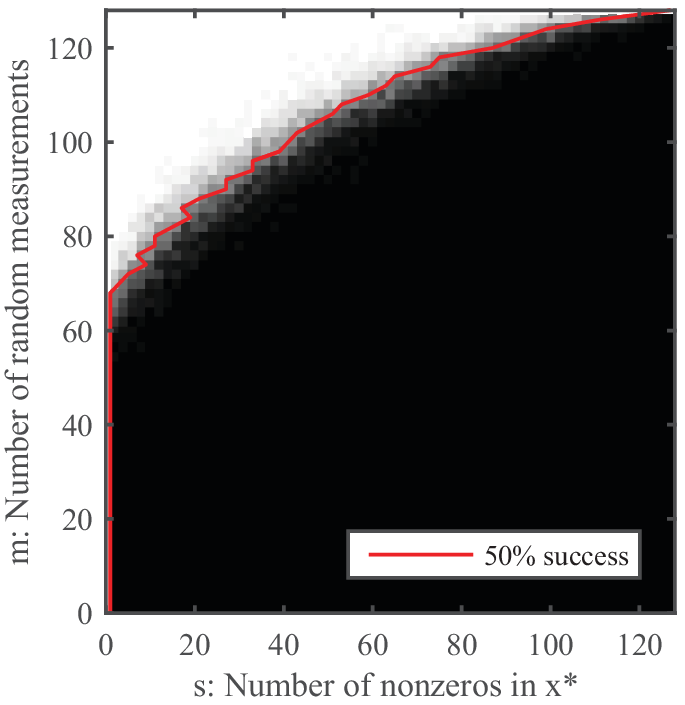}
\label{fig_fourth_case}}
\hfil
\subfloat[]{\includegraphics[width=1.10in]{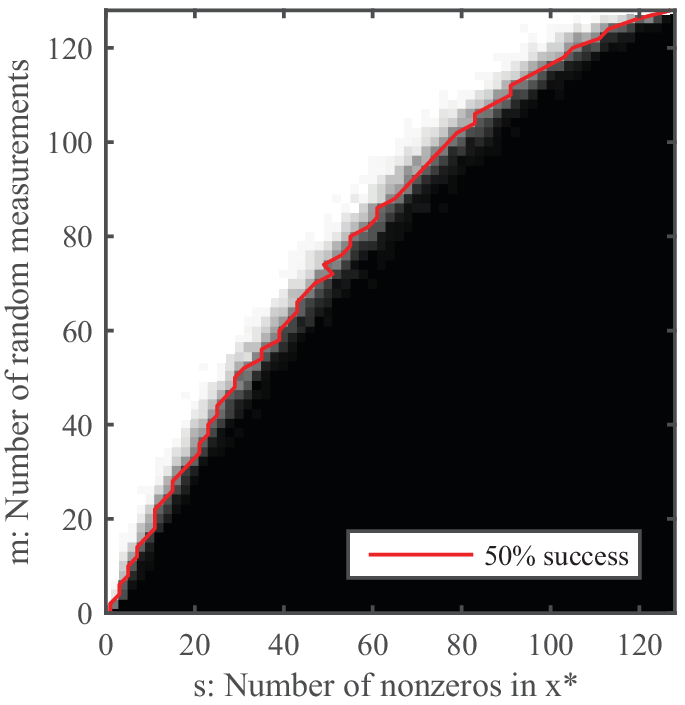}
\label{fig_fifth_case}}
\hfil
\subfloat[]{\includegraphics[width=1.10in]{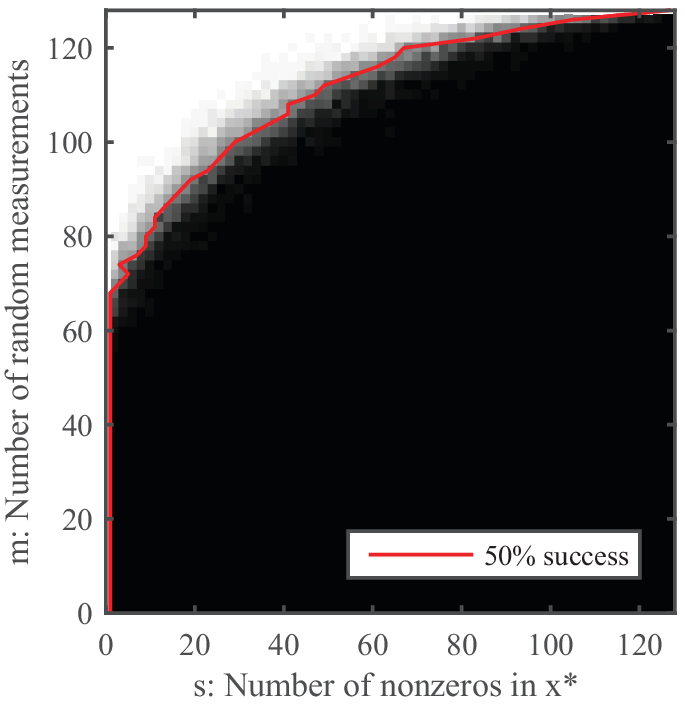}
\label{fig_sixth_case}}
\caption{The phase transition curves for different shifts: (a) $\lambda \vphi=\bm{0}$; (b) $\lambda \vphi= \frac{\text{sign}(\vx^\star)}{2}$; (c) $\lambda \vphi=-\frac{\text{sign}(\vx^\star)}{2}$; (d) $(\lambda \vphi)_{I}=\bm{0}$ and $(\lambda \vphi)_{I^c}=\bm{1}$; (e) $(\lambda \vphi)_{I}=\frac{\text{sign}(\vx^\star_I)}{2}$ and $(\lambda \vphi)_{I^c}=\bm{0}$ except an arbitrary $i \in I^c$ satisfying $\lambda \vphi_i=1/4$; (f) $(\lambda \vphi)_{I}=-\frac{\text{sign}(\vx^\star_I)}{2}$ and $(\lambda \vphi)_{I^c}=\bm{1}$. The brightness of each point reflects the observed probability of success, ranging from black(0\%) to white(100\%).}
\label{fig: PhaseTransition}
\end{figure*}

\subsection{Geometrical Interpretation}
In this subsection, we will present a geometrical interpretation for our main results.

Theorem \ref{thm: Generalcase} reveals that the number of measurements required for successful reconstruction is determined by the spherical Gaussian width of the tangent cone of $f$ at $\vx^\star$, i.e., $w(\mathcal{T}_f \cap S^{n-1})$. Recall that the normal cone $\mathcal{N}_f$ of $f$ at $\vx^\star$ is the polar of its tangent cone $\mathcal{T}_f$. This implies that the larger the normal cone $\mathcal{N}_f$, the less the number of measurements required for successful recovery.

In the classical CS problem \eqref{eq: Classical_Problem}, the subdifferential of the objective $\norm{\vx}_1$ at $\vx^\star$ is $\partial \norm{\vx^\star}_1 $. If $\bm{0} \notin \partial \norm{\vx^\star}_1$, then the corresponding normal cone is $\mathcal{N}_0 = \text{cone}\{\partial \norm{\vx^\star}_1\}$. For the optimization problem (\ref{eq: Main_Problem}), the subdifferential of the objective $f(\vx) = \norm{\vx}_1 - \lambda \langle \vx, \vphi\rangle$ at $\vx^\star$ is $\partial \norm{\vx^\star}_1 - \lambda \vphi$, which is a shifted version of $\partial \norm{\vx^\star}_1 $. If $\bm{0} \notin \partial \norm{\vx^\star}_1 - \lambda \vphi$, then $\mathcal{N}_f= \text{cone}\{\partial \norm{\vx^\star}_1 - \lambda \vphi\}$. In order to achieve a better performance for \eqref{eq: Main_Problem}, it is required that the prior information is good enough such that $\mathcal{N}_f$ is larger than $\mathcal{N}_0$. This will lead to a smaller tangent cone and hence less number of measurements required for successful recovery.

For convenience, we consider three different kinds of shifts (or prior information) for $\partial \norm{\vx^\star}_1 $, namely,
\begin{itemize}
\item a shift on the support set of $\vx^\star$ if $\lambda \vphi$ satisfies
\begin{equation*}
    \left\{
    {\begin{array}{ll}
     \lambda \vphi_i \neq 0, & \exists~ i \in I,  \\
     \lambda \vphi_i   =  0, & \forall~ i \in I^c.
   \end{array} }
    \right.
\end{equation*}
\item a shift on the complement of the support set of $\vx^\star$ if $\lambda \vphi$ satisfies
\begin{equation*}
    \left\{
    {\begin{array}{ll}
     \lambda \vphi_i   = 0, & \forall~  i \in I,  \\
     \lambda \vphi_i  \neq  0, & \exists~  i \in I^c.
   \end{array} }
    \right.
\end{equation*}
\item an arbitrary shift if $\lambda \vphi$ satisfies
\begin{equation*}
    \left\{
    {\begin{array}{ll}
     \lambda \vphi_i \neq  0, & \exists~  i \in I,  \\
     \lambda \vphi_i \neq  0, & \exists~  i \in I^c.
   \end{array} }
    \right.
\end{equation*}
\end{itemize}
Clearly, different shifts have different effects on the performance of optimization problem \eqref{eq: Main_Problem}. To illustrate this, we consider the two dimensional case. We set $\vx^\star=[1,0]^T$ and consider five different kinds of prior information: $\lambda_a\vphi_a=\[0.5,0\]^T,~\lambda_b\vphi_b=\[-0.5,0\]^T,~\lambda_c\vphi_c=\[0,-1\]^T,~\lambda_d\vphi_d=\[0.5,-0.2\]^T,$ and $\lambda_e\vphi_e=\[0.5,-1\]^T$. The results are shown in Fig. \ref{fig: GeometricalAnalysis}.

The shifts in Fig. \ref{fig: GeometricalAnalysis}(a) and \ref{fig: GeometricalAnalysis}(b) are shifts on the support of $\vx^\star$. In Fig. \ref{fig: GeometricalAnalysis}(a), the subdifferential moves toward the original, which enlarges the normal cone and hence decreases the number of measurements. The opposite result is shown in Fig. \ref{fig: GeometricalAnalysis}(b). The shift in Fig. \ref{fig: GeometricalAnalysis}(c) is a shift on the complement of the support of $\vx^\star$, which leads to a growth of sample size because the normal cone is narrowed. The shifts in Fig. \ref{fig: GeometricalAnalysis}(d) and in Fig. \ref{fig: GeometricalAnalysis}(e) are arbitrary shifts. We can know from Fig. \ref{fig: GeometricalAnalysis}(d) and \ref{fig: GeometricalAnalysis}(e) that different arbitrary shifts may lead to different results: the number of measurements of Fig. \ref{fig: GeometricalAnalysis}(d) gets reduced  while that of Fig. \ref{fig: GeometricalAnalysis}(e) gets increased.

These results can be theoretically explained by Theorem \ref{thm: SpecialCase}. Indeed, direct calculation of $v$ \eqref{parameter_v} in different cases leads to
$$v_0=2,v_a=1.25, v_b=3.25, v_c=5, v_d=1.69,\text{and}~v_e=4.25,$$
where $v_0$ corresponds to the classical CS case, and $v_a$-$v_e$ represents the results of Fig. \ref{fig: GeometricalAnalysis}(a)-(e) respectively. It is not hard to find that
the less $v$, the better the performance.

\begin{figure*}[!t]
\centering
\subfloat[]{\includegraphics[width=1.10in]{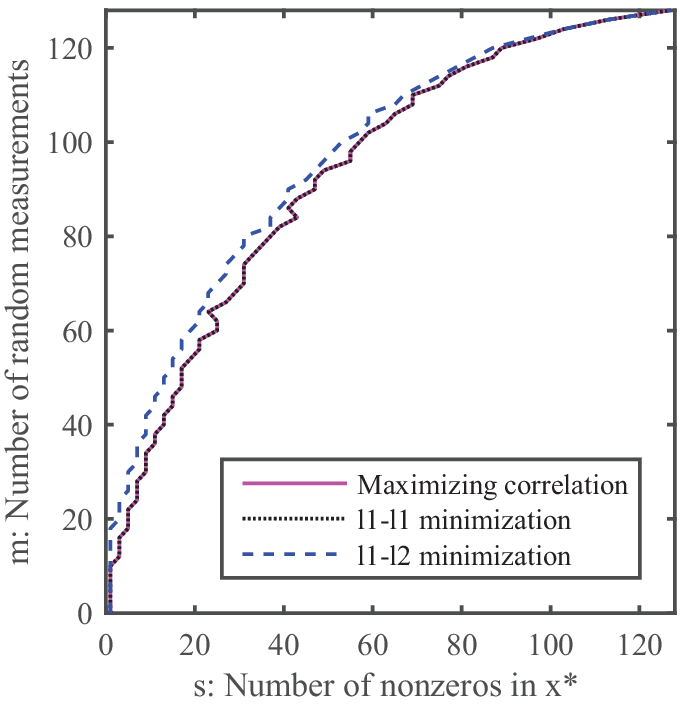}
\label{fig3_first_case}}
\hfil
\subfloat[]{\includegraphics[width=1.10in]{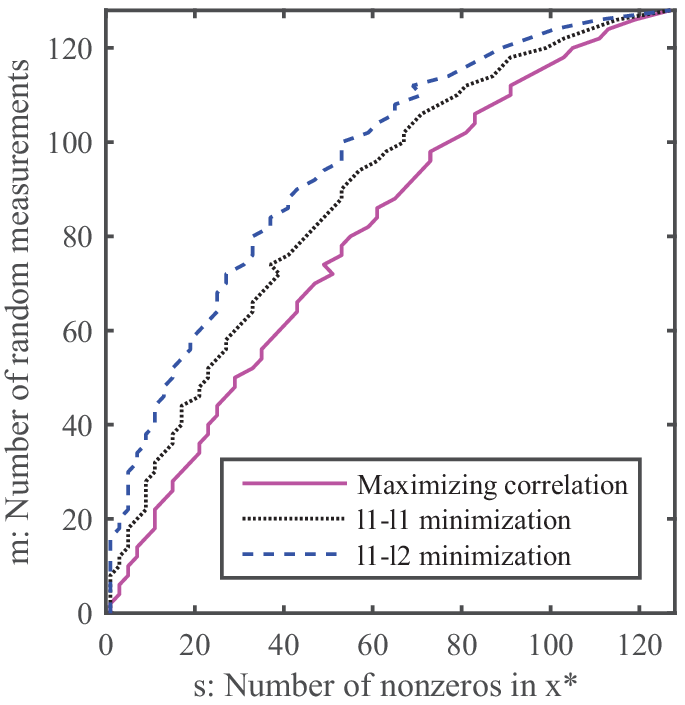}
\label{fig3_second_case}}
\hfil
\subfloat[]{\includegraphics[width=1.10in]{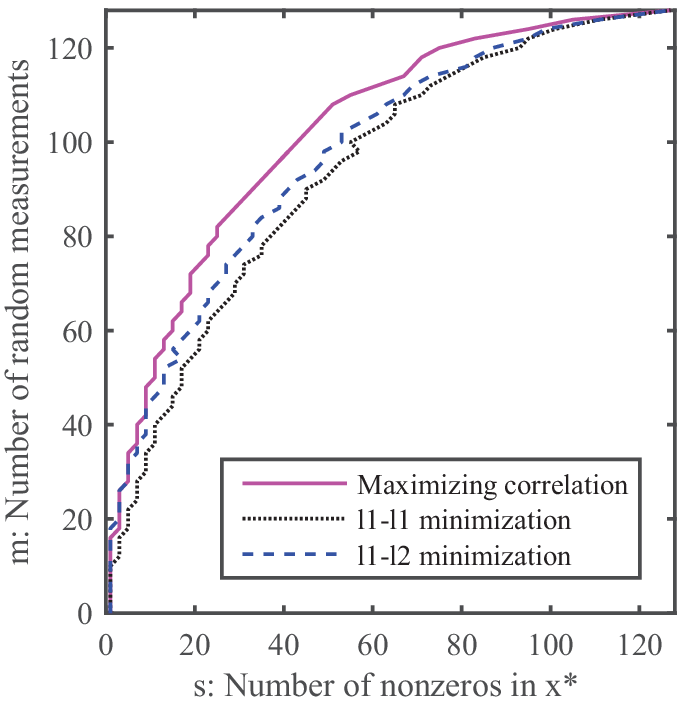}
\label{fig3_third_case}}
\hfil
\subfloat[]{\includegraphics[width=1.10in]{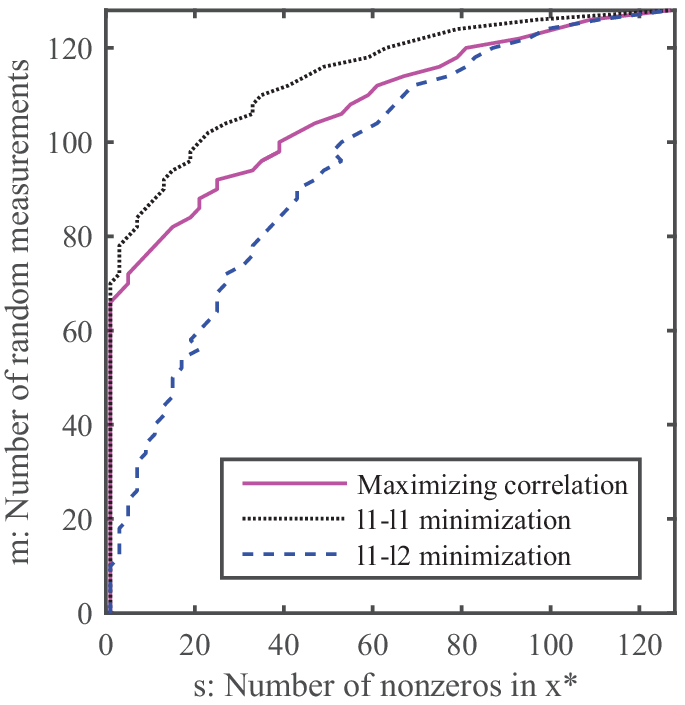}
\label{fig3_fourth_case}}
\hfil
\subfloat[]{\includegraphics[width=1.10in]{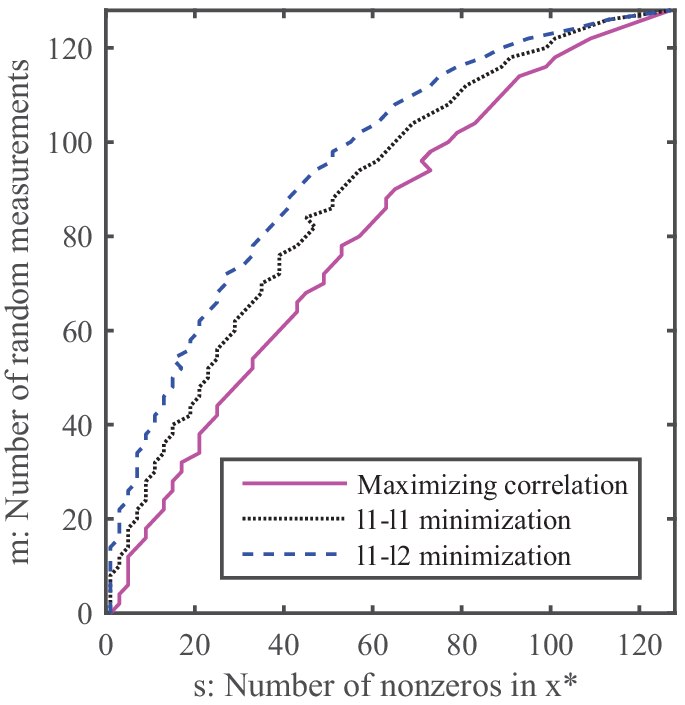}
\label{fig3_fifth_case}}
\hfil
\subfloat[]{\includegraphics[width=1.10in]{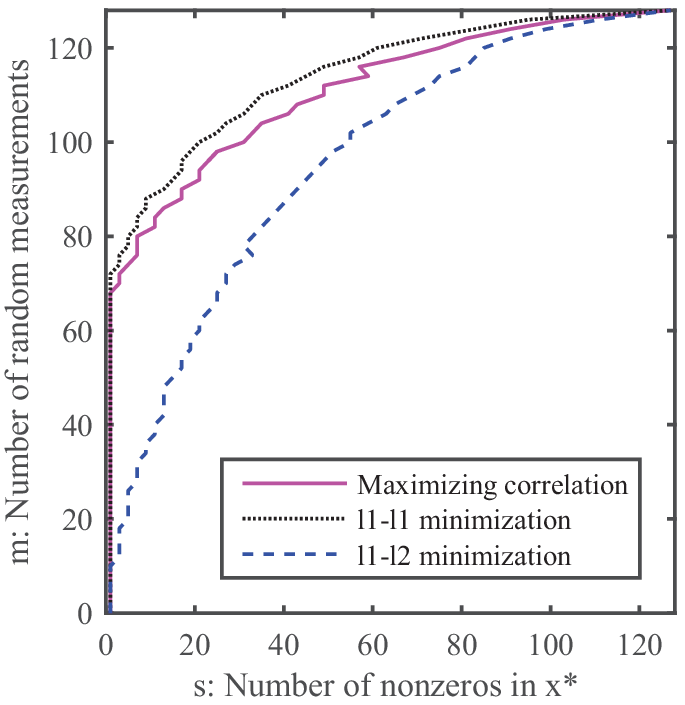}
\label{fig3_sixth_case}}
\caption{The phase transition curves of 50\% success for different shifts under three different methods, that is, CS with Maximizing Correlation, $\ell_1$-$\ell_1$ minimization and $\ell_1$-$\ell_2$ minimization. The shifts are (a) $\lambda \vphi=\bm{0}$; (b) $\lambda \vphi= \frac{\text{sign}(\vx^\star)}{2}$; (c) $\lambda \vphi=-\frac{\text{sign}(\vx^\star)}{2}$; (d) $(\lambda \vphi)_{I}=\bm{0}$ and $(\lambda \vphi)_{I^c}=\bm{1}$; (e) $(\lambda \vphi)_{I}=\frac{\text{sign}(\vx^\star_I)}{2}$ and $(\lambda \vphi)_{I^c}=\bm{0}$ except an arbitrary $i \in I^c$ satisfying $\lambda \vphi_i=1/4$; (f) $(\lambda \vphi)_{I}=-\frac{\text{sign}(\vx^\star_I)}{2}$ and $(\lambda \vphi)_{I^c}=\bm{1}$.}
\label{fig: PhaseTransitionofThree}
\end{figure*}

\section{Numerical Simulations}
In this section, we carry out some numerical simulations to verify the correctness of our theoretical results and compare the performance with previous methods. In these experiments, we draw the phase transition curves for different kinds of prior information: no shift, shifts on the support, shifts on the complement of the support, and arbitrary shifts. The original signal $\vx^\star \in \R^n$ is an $s$-sparse random vector and the measurement matrix $\vA \in \R^{m \times n}$ is a random symmetric Bernoulli matrix. Let $I$ be the set including all the locations of nonzeros in $\vx^\star$ and $\hat{\vx}$ be the estimator.  We set $n=128$ and $tol=10^{-2}$ for all the experiments. For a particular pair of $s$ and $m$, we make 50 trials, count the number of trials which succeed to recover $\vx^\star$, and calculate related probability. If the optimum of a trial satisfies
$$
\frac{\norm{\vx^\star - \hat{\vx}}_2}{\norm{\vx^\star}_2} < tol,
$$
we claim it as a successful trial. Let $m$ and $s$ increase from 0 to $n$ with step 2 respectively, then we can get a phase transition curve.

We consider six cases:
\begin{enumerate}
  \item[(a)] $\lambda \vphi=\bm{0}$. This is the classical CS model and $v_a=n$.
  \item[(b)] $\lambda \vphi= \text{sign}(\vx^\star)/2$. This is a shift on the support and  $v_b=n-3s/4$.
  \item[(c)] $\lambda \vphi=-\text{sign}(\vx^\star)/2$. This is a shift on the support and $v_c=n+5s/4$.
  \item[(d)] $(\lambda \vphi)_{I}=\bm{0}$ and $(\lambda \vphi)_{I^c}=\bm{1}$. This is a shift on the complement of the support and $v_d=4n-3s$. Here $\bm{1}$ denotes a vector whose entries are 1.
  \item[(e)] $(\lambda \vphi)_{I}=\text{sign}(\vx^\star_{I})/2$ and $(\lambda \vphi)_{I^c}=\bm{0}$ except an arbitrary $i \in I^c$ satisfying $\lambda \vphi_i=1/4$. This is an arbitrary shift and $v_e=n-3s/4+9/16$.
  \item[(f)] $(\lambda \vphi)_{I}=-\text{sign}(\vx^\star_{I})/2$ and $(\lambda \vphi)_{I^c}=\bm{1}$. This is an arbitrary shift and $v_f=4n-7s/4$.
\end{enumerate}

In Fig. \ref{fig: PhaseTransition}, we draw the phase transition curves in the above cases for the proposed approach. For shifts on the support, Fig. \ref{fig: PhaseTransition}(b) shows an improved performance while Fig. \ref{fig: PhaseTransition}(c) presents a deteriorative performance in contrast to the standard CS result in Fig. \ref{fig: PhaseTransition}(a). Comparing Fig. \ref{fig: PhaseTransition}(d) with Fig. \ref{fig: PhaseTransition}(a), we realize that the shift on the complement of the support makes the number of measurements increase whatever the sparsity is. In Fig. \ref{fig: PhaseTransition}(e), the simulation result shows that this arbitrary shift improves the performance a lot compared with Fig. \ref{fig: PhaseTransition}(a). However, Fig. \ref{fig: PhaseTransition}(f) presents an opposite result for the other arbitrary shift. All of these experiments coincide with the theoretical results of Theorem \ref{thm: SpecialCase} by comparing $v$ for six cases.

In Fig. \ref{fig: PhaseTransitionofThree}, we compare the performance of three methods (Maximizing Correlation method, $\ell_1$-$\ell_1$ minimization, and $\ell_1$-$\ell_2$ minimization) under different prior information.
In Fig. \ref{fig: PhaseTransitionofThree}(a), there is no prior information, so both Maximizing Correlation method and $\ell_1$-$\ell_1$ minimization reduce to Lasso. The result indicates that Lasso outperforms $\ell_1$-$\ell_2$ minimization in this setting. In Fig. \ref{fig: PhaseTransitionofThree}(b) and \ref{fig: PhaseTransitionofThree}(e), Maximizing Correlation method shows the best performance, followed by $\ell_1$-$\ell_1$ minimization and $\ell_1$-$\ell_2$ minimization. Fig. \ref{fig: PhaseTransitionofThree}(c) shows that $\ell_1$-$\ell_1$ minimization has the best performance, while Maximizing Correlation method has the worst performance. This is because $\vx^\star-\vphi$ is as sparse as $\vx^\star$, so $\ell_1$-$\ell_1$ minimization has a good performance. In contrast,  since $\vx^\star-\vphi$ is not sparse in Fig. \ref{fig: PhaseTransitionofThree}(d) and \ref{fig: PhaseTransitionofThree}(f), the performance is different: $\ell_1$-$\ell_2$ minimization becomes the best approach, followed by Maximizing Correlation method and $\ell_1$-$\ell_1$ minimization. In conclusion, the performance of these methods depends on the prior information and Maximizing Correlation method can achieve the best performance under certain prior information.

\section{conclusion} \label{sec: Conclusion}

This paper has proposed a new approach named Maximizing Correlation for CS with prior information. Theoretical analysis for this method has been established under sub-Gaussian measurements. Numerical simulations have been given to demonstrate the validity of the proposed approach. For future work, it would be of great practical interest to extend Maximizing Correlation approach from sparse vectors to other simple structures such as low rank matrices, block sparse vectors and so on.

%Keystones for our future work are as follows
%
%\emph{A tighter bound of measurements for robust reconstruction: }  There are two kinds of methods calculating the spherical gaussian width of tangent cone for $\ell_1$-norm , which refers to \cite[equation (9)]{foygel2014corrupted} and \cite[Proposition 3.10]{chandrasekaran2012convex}. Our analysis of $w(\mathcal{T}_f \cap S^{n-1})$ in Lemma \ref{lm: CalculateGaussianWidth} is based the former one. Calculating the other bound based on the latter one and combining them together may achieve a tighter bound of measurements for robust reconstruction.
%
%\emph{Extension to other simple structures:} It would be of great practical interest to extend Maximizing Correlation approach from sparse vectors to other simple structures such as low rank matrices, block sparse vectors and so on.

\ifCLASSOPTIONcaptionsoff
  \newpage
\fi

\end{document}